\documentclass[iop,twocolumn]{aastex63}
\hypersetup{citecolor=blue,urlcolor=blue}
\shorttitle{First Detection of Radio Emission from the Intermediate Mass Black Hole in POX 52}
\usepackage{multirow}
\usepackage{hyperref}
\usepackage{amssymb}
\usepackage{url}
\usepackage{subfiles}
\usepackage{mathrsfs}
\usepackage{helvet}
\makeatletter

\newcommand{\Rmnum}[1]{\expandafter\@slowromancap\romannumeral #1@}

\makeatother

\shortauthors{Yuan et al.}

\begin{document}

\title{First Detection of Radio Emission from the Intermediate Mass Black Hole in POX~52:\\ Deep Multi-Band Observations with ATCA and VLA}

\correspondingauthor{Qi Yuan, Hengxiao Guo, Minfeng Gu}
\email{Emails: yuanqi@cho.ac.cn (QY)\\
 hengxiaoguo@gmail.com (HXG)\\
 gumf@shao.ac.cn (MFG)} 

\author[0000-0003-4671-1740]{Qi Yuan}
\affiliation{Shanghai Astronomical Observatory, Chinese Academy of Sciences, 80 Nandan Road, Shanghai 200030, People's Republic of China}
\affiliation{Changchun Observatory, National Astronomical Observatories, Chinese Academy of Sciences, Changchun 130117, People's Republic of China}

\author[0000-0001-8416-7059]{Hengxiao Guo}
\affiliation{Shanghai Astronomical Observatory, Chinese Academy of Sciences, 80 Nandan Road, Shanghai 200030, People's Republic of China}

\author[0000-0002-4455-6946]{Minfeng Gu}
\affiliation{Shanghai Astronomical Observatory, Chinese Academy of Sciences, 80 Nandan Road, Shanghai 200030, People's Republic of China}

\author[0000-0002-5841-3348]{Jamie Stevens}
\affiliation{CSIRO Astronomy and Space Science, PO Box 76, Epping, NSW, 1710, Australia}

\author[0000-0002-8186-4753]{Philip G. Edwards} 
\affiliation{CSIRO Astronomy and Space Science, PO Box 76, Epping, NSW, 1710, Australia}

\author[0000-0001-5650-6770]{Yongjun Chen}
\affiliation{Shanghai Astronomical Observatory, Chinese Academy of Sciences, 80 Nandan Road, Shanghai 200030, People's Republic of China}

\author[0000-0002-4521-6281]{Wenwen Zuo}
\affiliation{Shanghai Astronomical Observatory, Chinese Academy of Sciences, 80 Nandan Road, Shanghai 200030, People's Republic of China}

\author[0000-0001-8416-7059]{Jingbo Sun}
\affiliation{Shanghai Astronomical Observatory, Chinese Academy of Sciences, 80 Nandan Road, Shanghai 200030, People's Republic of China}
\affiliation{University of Chinese Academy of Sciences, 19A Yuquan Road, 100049, Beijing, People's Republic of China}

\author[0000-0002-2322-5232]{Jun Yang}
\affiliation{Department of Space, Earth and Environment, Chalmers University of Technology, Onsala Space Observatory, SE-43992 Onsala, Sweden}

\author[0000-0003-1523-9164]{Paulina Lira}
\affiliation{Departamento de Astronom\`{\i}a, Universidad de Chile, Casilla 36D, Santiago, Chile}

\author[0000-0003-4341-0029]{Tao An}
\affiliation{Shanghai Astronomical Observatory, Chinese Academy of Sciences, 80 Nandan Road, Shanghai 200030, People's Republic of China} 

\author[0000-0002-2432-2587]{Renzhi Su}
\affiliation{Research Center for Astronomical Computing, Zhejiang Laboratory, Hangzhou 311100, People's Republic of China}

\author[0000-0001-9321-6000]{Yuanqi Liu}
\affiliation{Shanghai Astronomical Observatory, Chinese Academy of Sciences, 80 Nandan Road, Shanghai 200030, People's Republic of China} 

\author[0000-0002-1010-7763]{Yijun Wang}
\affiliation{School of Astronomy and Space Science, Nanjing University, 163 Xianlin Avenue, Nanjing 210023, People's Republic of China}
\affiliation{Key Laboratory of Modern Astronomy and Astrophysics, Nanjing University, Ministry of Education, 163 Xianlin Avenue, Nanjing 210023, People's Republic of China}

\author[0000-0002-8684-7303]{Ning Chang}
\affiliation{Xinjiang Astronomical Observatory, Chinese Academy of Sciences, 150 Science 1-Street, Urumqi 830011, People's Republic of China}

\author[0000-0003-3166-5657]{Pengfei Jiang}
\affiliation{Xinjiang Astronomical Observatory, Chinese Academy of Sciences, 150 Science 1-Street, Urumqi 830011, People's Republic of China}

\author[0000-0002-8315-2848]{Ming Zhang}
\affiliation{Xinjiang Astronomical Observatory, Chinese Academy of Sciences, 150 Science 1-Street, Urumqi 830011, People's Republic of China}
\affiliation{Key Laboratory of Radio Astronomy, CAS, 150 Science 1-Street, Urumqi 830011, People's Republic of China}
\affiliation{Xinjiang Key Laboratory of Radio Astrophysics, 150 Science 1-Street, Urumqi 830011, People's Republic of China}

\begin{abstract}
We present the first multi-band centimeter detection of POX 52, a nearby dwarf galaxy believed to harbor a robust intermediate mass black hole~(IMBH). We conducted the deep observations using the Australia Telescope Compact Array (ATCA), spanning frequencies from 4.5 to 10~GHz, as well as the sensitive observations from the Karl G. Jansky Very Large Array (VLA) operating in its most extended A-configuration at S~band (2--4~GHz) and C~band (4--8~GHz). In the ATCA observations, the source shows a compact morphology, with only one direction marginally resolved. The higher resolution of the VLA allowed us to slightly resolve the source, fitting it well with a two-dimensional Gaussian model. The detected radio emission confirms the presence of Active Galactic Nucleus (AGN) activity, indicating either a low-power jet or AGN-driven winds/outflows. Our dual-epoch observations with ATCA and VLA, together with previous non-detection flux density upper limits, reveal radio emission variability spanning two decades. In addition, we find that POX 52 aligns well with the low-mass extension of the fundamental plane for high-accretion, radio-quiet massive AGNs. 
\end{abstract}


\section{introduction}
Intermediate mass black holes (IMBHs), with masses of $10^2 - 10^6 M_{\odot}$~\citep{Mezcua.17.ijmpd}, bridge the gap between stellar-mass black holes (BHs) and supermassive black holes (SMBHs, $M_{\rm BH} \ge 10^6 M_{\odot}$). As observations from the James Webb Space Telescope (JWST) push the discovery of first-generation quasars to earlier and earlier cosmic epochs \citep[e.g.,][]{Kokorev23}, understanding how SMBHs formed has become an urgent problem in cosmology \citep{Valiante18}. Detecting populations of IMBHs could provide constraints on these elusive black hole seeds.
However, probing these seed black holes in the early Universe is beyond the reach of current observational capabilities. 

Additionally, rapidly growing SMBHs at high redshifts quickly lose all traces of their origins. Fortunately, theoretical work suggests that observations of local, leftover IMBHs in dwarf galaxies may also provide crucial clues to understanding seeding processes, as these galaxies are relatively pristine and have undergone fewer mergers and less accretion since their formation compared to their supermassive counterparts \citep{Volonteri08,vanWassenhove10,Inayoshi.20.araa}. Therefore, confirming the IMBH nature serves as a first step in understanding SMBH formation history.

About two decades ago, NGC 4395 and POX 52 were considered two prototypical IMBH hosts in the local Universe, supported by multi-wavelength evidence including optical spectroscopic signatures \citep{Filippenko.03.apjl,barth.04.apj}. With advancements in observational technology, the number of IMBH candidates has increased to hundreds. A growing body of research leverages multi-wavelength tracers, including optical spectroscopic diagnostics~\citep{green.04.apj, dong.12.apj, Reines.13.apj, liu.18.apjs, salehirad.22.apj}, optical/X-ray variability~\citep{Baldassare.18.apj, Palomera.20.apj,Kamizasa12}, coronal lines \citep{Wasleske24}, mid-infrared color diagnostics~\citep{Satyapal.14.apj, Sartori.15.mn}, and radio/X-ray emissions~\citep{Reines.20.apj,Sargent.22.apj}, to explore active IMBHs within low-mass/dwarf galaxies. It is important to note that selection biases exist among IMBH candidates identified across different bands~\citep{Askar.23.arxiv, Wasleske24}. Thus, insightful multi-band follow-up studies can help confirm their IMBH nature and enhance our understanding of the physics and characteristics of this population.

The two most reliable IMBH candidates, in the dwarf galaxies NGC 4395 and POX 52, exhibit very different host properties and accretion modes.
NGC 4395 is a late-type, bulgeless spiral dwarf galaxy at z = 0.00106 ($\sim$ 4.5 Mpc).
Reverberation mapping of its H$\alpha$ line indicates a black mass mass of $\sim$ 1.7 $\times$ 10$^4$ $M_{\odot}$, with an Eddington ratio around 0.06 \citep{Woo19,Cho21,Pandey24}. 
In contrast, POX~52 (J1202$-$2056, at $z$ = 0.021, or $\sim$~90 Mpc) is a dwarf elliptical galaxy lacking any obvious spiral, disk-like structure, or clumps,  
and accreting close to the Eddington limit~\citep{barth.04.apj}. The estimated single-epoch black hole mass is $\sim$ 1.6 $\times$ $10^5\ M_{\odot}$, based on Keck spectroscopic observations \citep{barth.04.apj}. POX 52 has been extensively studied in multi-band observations \citep{barth.04.apj,thornton.08.apj, Kawamuro.24.apj} and exhibits significant optical variability and dust echo, offering potential for accurate black hole mass estimation through reverberation mapping techniques (Sun et al., in prep.).

Radio properties are crucial for studying black holes and can provide independent evidence of AGN activity, as well as insights into the characteristics of IMBHs. In the radio regime, NGC 4395 has been observed across various frequencies using interferometer arrays with different angular resolutions to investigate its radio origins~\citep {Saikia18, yangjun.22.mn, Nandi.23.apj}. Research has progressed from initial explorations of radiation properties on the arcsec scale to sub-arcsec detections, which could determine whether there is a sub-parsec-scale continuous or episodic jet tracing the accreting IMBH~\citep[see][and references therein]{yangjun.22.mn, yangjun.23.mn}. In contrast, no significant radio emission from POX 52 have been detected before this work. Previous radio observations employed the historical Very Large Array~\citep[VLA,][]{thompson.80.apjs} in its A configuration with C band~\citep{greene.06.apj}~(sensitivity: 26~$\mu$Jy/beam) and VLA Sky Survey~\citep[VLASS][]{lacy.20.pasp} Epoch~1, 2 and 3 Quick Look~\citep{gordon.21.apjs}~(sensitivities: 155, 130, and 140~$\mu$Jy/beam, respectively). The failure to detect radio emission from POX 52 may be attributed to sensitivity limitations, its inherently faint nature—potentially linked to the anti-correlation between accretion rate and relative radio strength compared to the optical \citep[see][]{Luis.02.apj}—or the possibility that previous observations performed during a radio quiescent state~\citep{green.06a.apj}. As one of the most promising IMBH candidates, and in contrast to the low accretion state of NGC 4395, it serves as a representative for high-accretion sources. Thus, high-sensitivity deep observations are essential.

This paper reports our first multi-band radio detections of POX 52 and characterizes its radio properties.
The layout is as follows. In \S \ref{sec:obs}, we describe our observations and data reduction. We present the result in \S \ref{sec:results} and discuss our findings in \S \ref{sec:diss}. We draw our conclusions in \S \ref{sec:con}. Throughout the paper, we use the cosmological parameters $\Omega_{\rm m}$ = 0.27, $\Omega_{\rm \Lambda}$ = 0.73, and $H_{\rm 0} =71 \mathrm{~km} \mathrm{~s}^{-1}\ \mathrm{Mpc}^{-1}$. In this adopted $\Lambda$CDM cosmology, the angular scale at the source redshift $z$ = 0.021 is 419 $\mathrm{pc}\ \mathrm{arcsec^{-1}}$.

\begin{deluxetable*}{lcccccccccc}
\scriptsize
\tablecaption{ATCA and VLA Observations}
\tablehead{
    \colhead{Telescope} &
    \colhead{Frequency} & 
    \colhead{$t_{\rm on\ source}$} &
    \colhead{Peak Intensity} & 
    \colhead{\shortstack{Integrated \\ Flux Density}} & 
    \colhead{Synthesized Beam} &
    \colhead{Size} & 
    \colhead{rms} & 
    \colhead{SNR} &
    \colhead{$\mathscr{F}$} \\
    & (GHz) & (hr) & ($\mu$Jy/beam) & ($\mu$Jy) & (arcsec, arcsec, deg) & (arcsec, arcsec) & ($\mu$Jy/beam) & & \\
    (1) & (2) & (3) & (4) & (5) & (6) & (7) & (8) & (9) & (10)
}
\startdata
ATCA & 4.5--6.5 & 9 & 153 $\pm$ 12 & 260 $\pm$ 31 & (3.08, 0.93, $-0.57$) & unresolved & 11.4 & 13 &  \\
ATCA & 8--10 & 9 & 95 $\pm$ 9 & 160 $\pm$ 22 & (2.01, 0.61, $-0.58$) & unresolved & 8.2 & 12 &  \\
VLA & 2--4 (S band) & 0.99 & 157 $\pm$ 5  & 229 $\pm$ 11 & (1.40, 0.69, $-25.15$) &  (0.82, 0.50) & 4.8 & 32 & 1.46 \\
VLA & 4--8 (C band) & 0.97 & 82 $\pm$ 3 & 146 $\pm$ 8 & (0.74, 0.35, $-26.16$) &(0.44, 0.41) &3.0 & 27 & 1.81 \\
\enddata
\tablecomments{
    Column~1: radio interferometer. The ATCA and VLA observations were performed on November 18, 2022 and July 28, 2023, respectively.
    Column~2: observing frequency.
    Column~3: integration time on source.
    Column~4: peak intensity.
    Column~5: integrated flux density.
    Column~6: synthesized beam.
    Column~7: the source sizes deconvolved from the beam.
    Column~8: root mean square noise level. 
    Column~9: signal to noise ratio (SNR). 
    Column~10: ratio between integrated flux and peak flux ($\mathscr{F}$).\\
    }
\label{table:info_obs}
\end{deluxetable*}

\section{Observations \& data reduction }\label{sec:obs}
\subsection{ATCA observations}
POX~52 was observed with the Australia Telescope Compact Array (ATCA) in its 4 cm band, using the Compact Array Broadband Backend (CABB) system~\citep{cabbpaper}. The observations were made in the 6~C configuration on 2022 November 18 (Program ID: C3510) and were carried out with two 2 GHz wide intermediate frequencies (IF) of 4.5--6.5 GHz (centered at 5.5 GHz) and 8--10 GHz (centered at 9~GHz). PKS B1934$-$638 was used as bandpass and flux density calibrator while 1143$-$245 was used as complex gain calibrator. The data reduction was carried out with the software {\tt Miriad}~\citep{miriad}, following standard procedure. 

Automatic Radio Frequency Interference (RFI) flagging was performed using the task \textit{pgflag} before calibration. In total, 12.7\% of the 5.5~GHz band was flagged as bad, and 18.5\% of the 9~GHz band. Standard calibration involved bandpass and flux density calibration on PKS B1934$-$638 using the {\tt Miriad} tasks \textit{mfcal} and then \textit{gpcal}, and this was applied to the gain calibrator 1143$-$245. Time-varying gains and polarization leakage calibration was done using the task \textit{gpcal} on 1143$-$245, and these gains were transferred and applied to the source before imaging.

Imaging was performed using the tasks \textit{invert}, \textit{mfclean} and \textit{restor}, to create two continuum images, one for each of the IFs, using multi-frequency synthesis. Images were made in Stokes I, with a Briggs visibility weighting robustness parameter of $-1$. Cleaning with \textit{mfclean} was done with both the Clark and Hogbom algorithms down to a depth of 40~$\mu$Jy in both IFs. Images were restored with a Gaussian synthesized beam of $3\farcs084\times0\farcs9318$
with position angle of $-0.574^{\circ}$ for the 5.5~GHz image, and 
$2\farcs013\times0\farcs6068$ with position angle of $-0.578^{\circ}$ for the 9~GHz image. The central 10\% of the 5.5~GHz image has an rms noise level of 11.4~$\mu$Jy, while the 9~GHz image has an rms noise level of 8.2~$\mu$Jy. Source flux density fits were made using the task \textit{imfit}.

\subsection{VLA observations}
Our target was also observed with the VLA in the most extended A configuration in 2023 July 28 (Program ID: 23A-163).
Continuum observations were carried out at $S$ band with 2$\times$1~GHz basebands centered at 3~GHz; C band with 2$\times$2~GHz basebands centered at 6~GHz. For better $uv$ coverage, we alternated between the two bands, repeating this sequence three times during the overall scheduling block (SB). The first two sub-SB began with observations of a flux calibrator (3C~286) and then iterated between observing a complex gain calibrator (PKS~J1159$-$2148) and the target source, ending with the observation of the former. The third sub-SB mirrored the mode of the preceding two but omitted the flux scale observation. The raw data were reduced with the Common Astronomy Software Applications package (\textsc{\footnotesize{CASA}}) version 6.5.4 ~\citep{McMullin.07.aspc} with additional manual flagging.

\begin{figure*}
    \includegraphics[width=8.35cm]{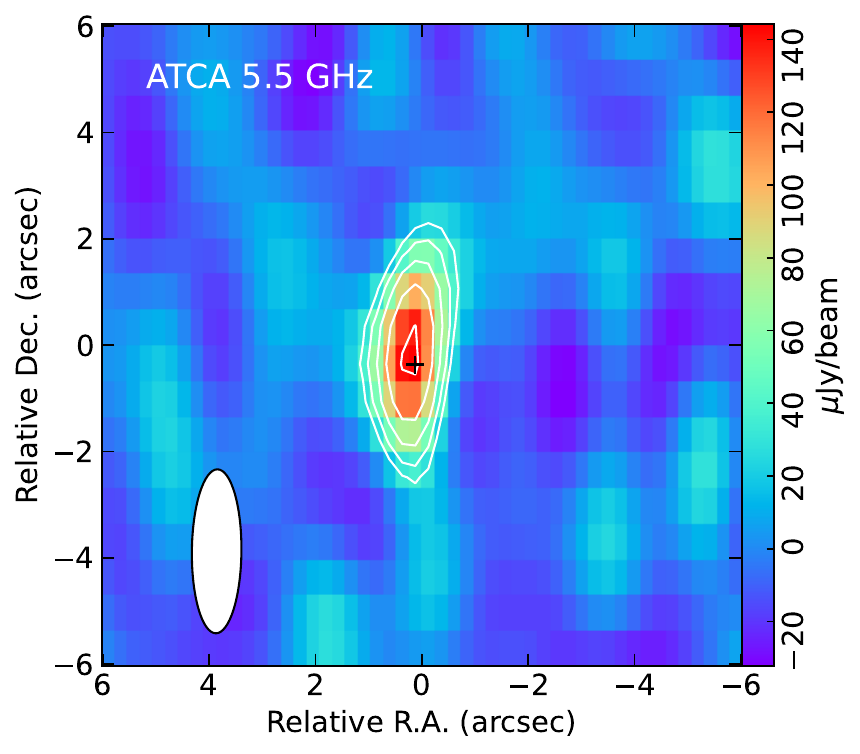}
    \includegraphics[width=8.5cm]{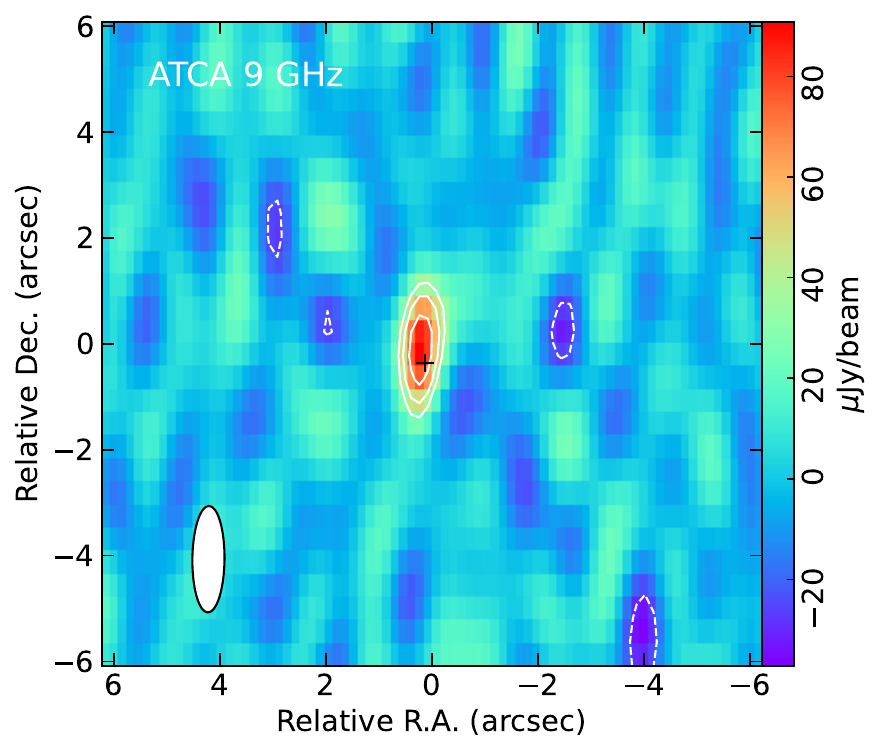}\\
    \includegraphics[width=8.5cm]{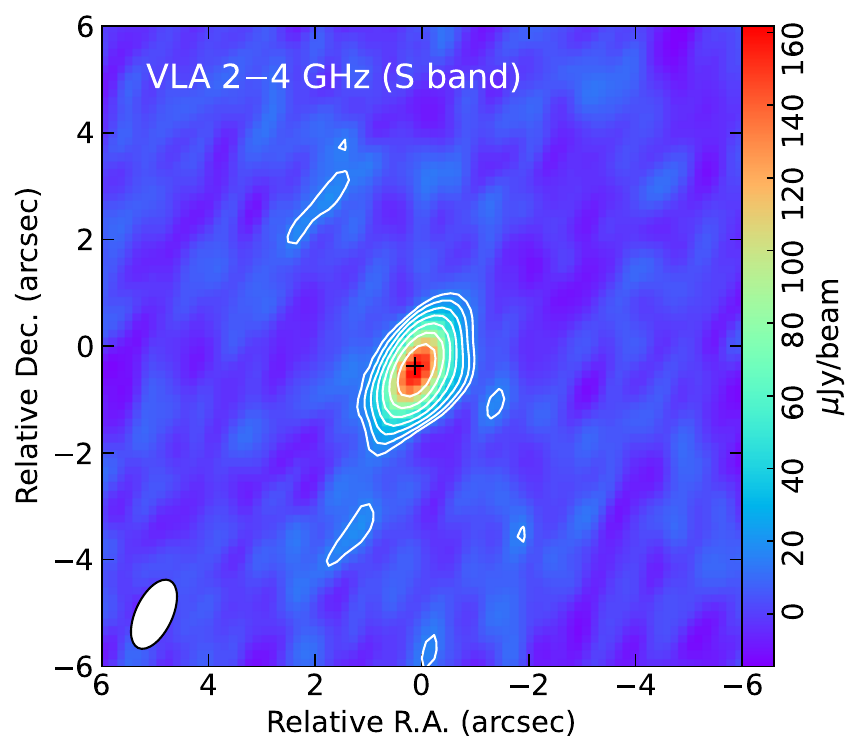}
    \includegraphics[width=8.5cm]{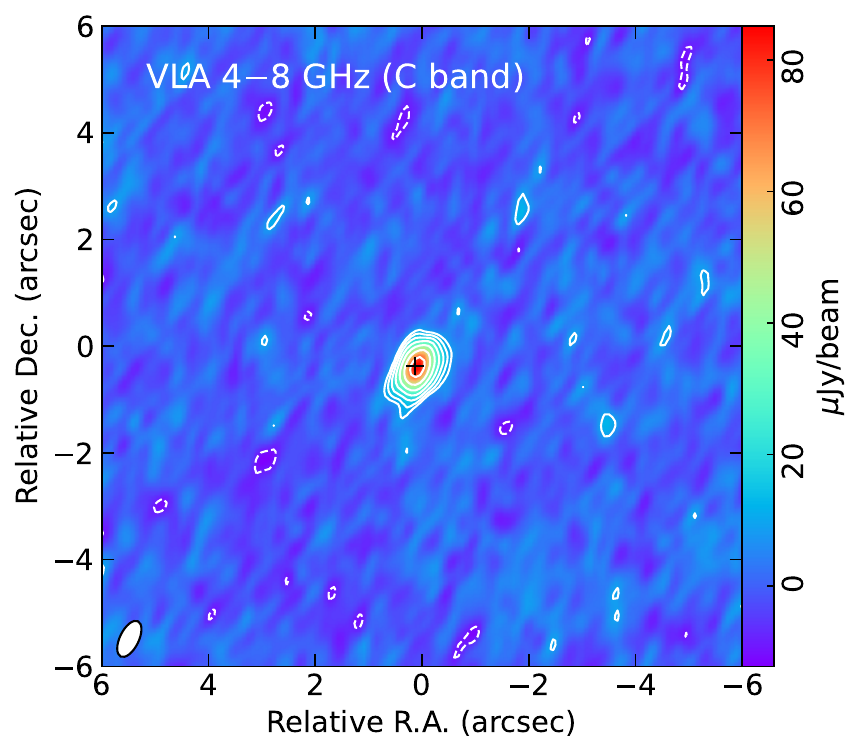}
    \caption{Naturally weighted \textsc{\footnotesize{CLEAN}} map of POX~52 observed by ATCA and VLA, centered at the radio positions obtained by model fitting. Contours in the map are plotted at multiples of $-1$, 1, $\sqrt{2}$, 2, 2$\sqrt{2}$, 4, 4$\sqrt{2}$, 8, 8$\sqrt{2}$, 16 $\times$~3$\sigma$, where $\sigma$ is the local rms noise. The off-source background noise of each observation is 11.4, 8.2, 4.8, 3.0 $\mu$Jy/beam, respectively. The white ellipses in the bottom left corner of each panel represent the full width at half-maximum (FWHM) of the restoring beam. The cross indicates the optical center obtained from Gaia Data Release 3 \citep[Gaia DR3,][]{gaia.23.summary}.}
    \label{fig:intensity}
\end{figure*}

The calibrated visibilities were imaged using the \textsc{\small{CLARK}} deconvolution algorithm~\citep{Clark.88.aa} \textsc{\footnotesize{CASA}} task \textsc{\small{TCLEAN}}. 
The \textsc{\footnotesize{CLEAN}} map, as shown in Figure~\ref{fig:intensity}, was reconstructed by adopting the natural weighting to optimize the sensitivity.
The background noise level is obtained by examining pixels well away from the source, executed by the task \textsc{\small{IMSTAT}}.
In addition to the maps with natural weighting, we also produced lower-resolution $uv$-tapered C-band 4--6 GHz and 6--8 GHz maps that match the resolution of the S band for spectral shape analysis only. It is implemented by splitting the total bandwidth of the C band into two portions and assigning lower weights to the visibility corresponding to small scales (i.e., long baselines). 

We utilized the \textsc{\footnotesize{CASA}} task \textsc{\small{IMFIT}} to acquire source parameters, including integrated flux density, peak intensity, and size. The final observation image in each band can be well-fitted by a single 2D elliptical Gaussian component. The source information and corresponding errors are listed in Table~\ref{table:info_obs}; the errors are derived from the uncertainty estimates in the \textsc{\small{IMFIT}} model fitting combined with the 3$\%$ limit in the flux calibration~\citep{Perley.17.apjs}.

\section{Results and Analysis}\label{sec:results}
\subsection{Radio emission and spectral index}
\begin{figure}
    \includegraphics[width=9cm]{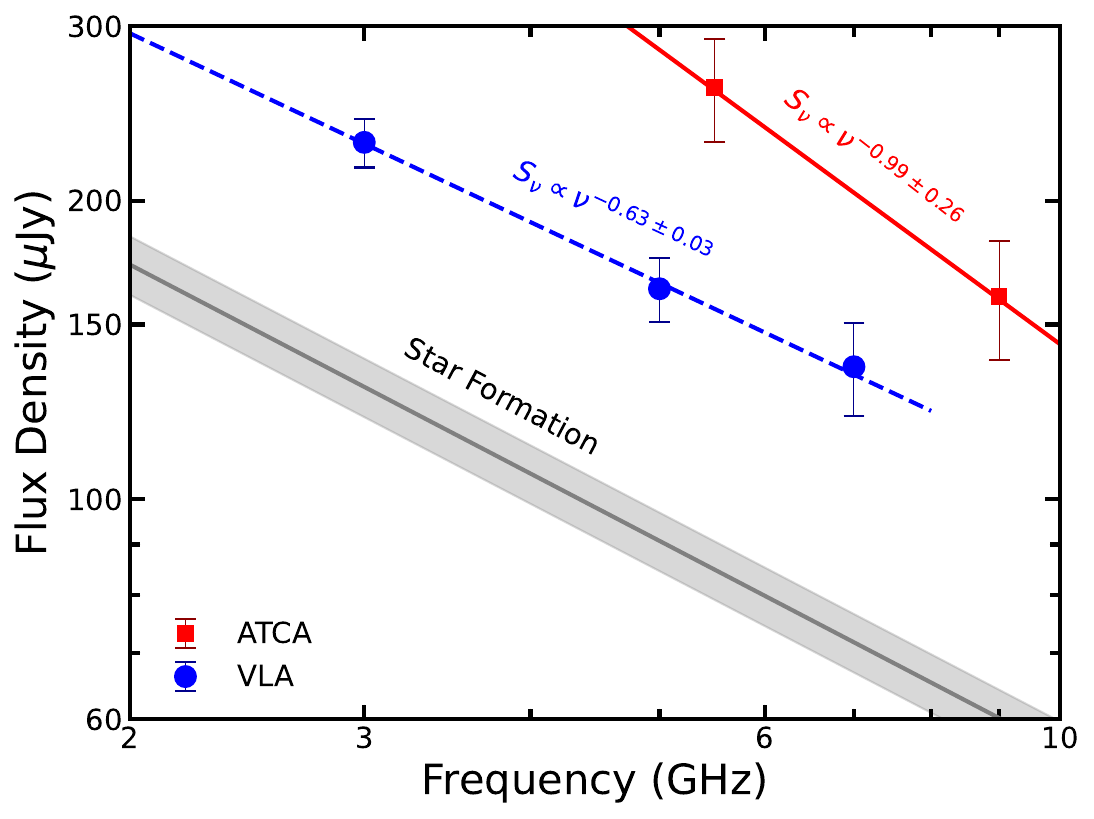}
    \caption{The radio spectra of POX~52.
    The red squares and blue dots represent ATCA and VLA observations, respectively. 
    The red solid and blue dashed lines show the best-fitting power-law spectra for ATCA and VLA observations. The gray solid line denotes the anticipated flux density values for star formation, while the surrounding gray shaded region reflects the associated uncertainty in the expected flux density due to the uncertainty in the SFR.
    }
    \label{fig:spectral}
\end{figure}

POX~52 is detected in both bands in ATCA and VLA observations with SNRs $\geq$ 10, respectively. The ATCA and VLA \textsc{\footnotesize{CLEAN}} maps are shown in Figure~\ref{fig:intensity}, and the radio emission is consistent with the optical center provided by Gaia~DR3~\citep{gaia.23.summary}. We calculated the spectral index ($\alpha$) which is defined as the power-law spectrum $S_{v} \propto \nu^{-\alpha}$, where $S_{\nu}$ is the integrated flux density at frequency $\nu$. 
Through a non-linear least-square minimization, we obtained a best-fit spectral index of 0.63, with an uncertainty of 0.03, for the VLA S band and $uv$-tapered C band observations. 
For ATCA observations, which include only two flux density values, the two-point spectral index $\alpha$ is transformed into $\alpha=\frac{\log \left(S_1 / S_2\right)}{\log \left(v_1 / v_2\right)}$ with its uncertainty estimated as
$\sqrt{\left(\sigma_{f_1} / S_{f_1}\right)^2+\left(\sigma_{f_2} / S_{f_2}\right)^2} / \ln \left(f_2 / f_1\right)$, where $\sigma_{f_{1,2}}$ and $S_{f_{1,2}}$ represent the uncertainties and the flux density at the two frequencies~\citep{Ho.01.apjs}. The spectral index derived from the ATCA observation is $0.99 \pm 0.26$.
Figure~\ref{fig:spectral} shows the radio spectra of POX~52.

According to the conventional classification of a steep spectrum ($\alpha \geq$ 0.5), a flat spectrum ($0 \leq \alpha<$ 0.5), and an inverted spectrum $\alpha<$ 0) \cite[e.g.,][]{Panessa.13.mn}, our target falls into the category of steep spectra both in ATCA and VLA observations though with different angular resolution.
This spectrum is significantly steeper than expected for thermal emission ($\alpha \sim 0.1$) or a self-absorbed jet base ($\alpha \sim 0$), indicating optically thin synchrotron emission.

\subsection{Radio loudness}

The classification of AGN into radio-quiet and radio-loud categories is used to characterize the strength of their radio emissions, reflecting the diverse underlying physical processes driving these emissions. The traditional radio-optical radio-loudness is defined as 
$\mathcal{R}_{\mathrm{O}}=L_{\mathrm{5GHz}} / L_{\rm 4400\, \text{\AA}}$~\citep{Kellermann.89.aj},
where $L_{\mathrm{5GHz}}$ and $L_{4400 \text{\AA}}$ are the radio and optical flux densities at 5~GHz and 4400\,\AA\, in the rest frame, respectively. 
The rest-frame $L_{4400 \text{\AA}}$ is obtained from the observed spectrum in \citet{barth.04.apj}. We revisited the VLA C-band observations, dividing them into two 2 GHz basebands (4–6 GHz and 6–8 GHz) to obtain an accurate flux density at 5 GHz. This yielded a radio loudness ($\mathcal{R}_{\mathrm{O}}$) estimate of 0.57 for this source.

Regarding the radio–X-ray definition, $\mathcal{R}_{\mathrm{X}} = L_{\mathrm{R}}/L_{\mathrm{X}}$, where $L_{\mathrm{R}}$ is the radio luminosity at 5~GHz, and $L_{\mathrm{X}}$ is the integrated luminosity between 2 and 10 keV \citep{Terashima.03.apj}. The X-ray luminosity of POX 52, obtained from \citet{Kawamuro.24.apj}, is $\sim$ $10^{42}$ erg s$^{-1}$. The radio luminosity at~5 GHz is estimated by
\begin{equation}
L_{\mathrm{R}} = \nu L_{\nu}=\nu f_\nu \frac{4 \pi D_{\rm L}^2}{(1+z)^{1+\alpha}},
\end{equation}
where $\alpha$ is the spectral index based on our VLA observation, $f_\nu$ is the integrated flux density at 5~GHz, and $D_{\rm L}$ is the luminosity distance.
The $L_{\mathrm{R}}$ at 5~GHz is estimated to be $10^{36.92}$ erg s$^{-1}$, resulting in $\mathcal{R}_{\mathrm{X}}=10^{-5.08}$.

According to the traditional radio-quiet criterion based on either radio-optical \citep[$\mathcal{R}_{\mathrm{O}} < 10$,][]{Visnovsky.92.apj, Kellermann.94.aj} or radio–X-ray \citep[$\mathcal{R}_{\mathrm{X}} < 10^{-2.755\pm 0.015}$,][]{Panessa.07.a} ratio, the source is thus classified as radio-quiet by both criteria.

\subsection{Brightness temperature}
The brightness temperature ($T_{\rm b}$) of the fitted Gaussian component can be calculated using the following formula~\citep{Kovalev.05.aj}:
\begin{equation}
T_{\rm b}=1.22 \times 10^{12} \frac{f(1+z)}{\theta_{\mathrm{maj}} \theta_{\min } \nu_{\mathrm{obs}}^2} \mathrm{~K},
\end{equation}
where $z$ is the source redshift, $f$ is the integrated flux density of the fitted Gaussian component in Jy, $\theta_{\min}$ and $\theta_{\mathrm{maj}}$ are the FWHM dimensions of the Gaussian in milli-arcsec, and $\nu_{\mathrm{obs}}$ is the observing frequency in GHz. We estimated $T_{\rm b}$ using the highest frequency sub-band ($6-8$ GHz) from the VLA C-band observation since it provides the highest spatial resolution in our observations. The estimated $T_{\rm b}$ at 7~GHz is $\sim 10^{2}$~K. Note that the brightness temperature estimation for this source only serves as a lower limit, given the measured angular diameters are below the nominal angular resolution.

\subsection{Radio morphology}

For the ATCA observations, we could not deconvolve the source from the synthesized beam, suggesting that the source might be marginally resolved along one direction. In contrast, the high-resolution VLA observations can reveal finer morphological structures on smaller angular scales. To determine whether our source is resolved, we calculate the ratio of integrated flux to peak flux ($\mathscr{F}$), which serves as a direct indicator of the radio source's extension~\citep{Prandoni.00.aaps, Bondi.03.aa}.
The $\mathscr{F}$ is characterised by the equation~\citep{Huynh.05.aj}:
\begin{equation}
\mathscr{F} = \frac{S_{\text {int }}}{S_{\text {peak }}}=\frac{\theta_{\text {maj }} \theta_{\text {min }}}{b_{\text {maj }} b_{\text {min }}},
\end{equation}
where $\theta_{\text {maj}}$ and $\theta_{\text {min }}$represent the FWHM of the Gaussian component fitted, while $b_{\text {maj }}$ and $b_{\text {min }}$ denote the FWHM of the synthesized beam. According to the criterion provided by~\citet{Prandoni.00.aaps}, a source is considered resolved if $\mathscr{F} > 1.05 + \frac{10}{(\mathrm{SNR})^{1.5}}$. For the VLA S and C bands, the $\mathscr{F}$ values of 1.46 and 1.81 exceed the predicted thresholds of 1.11 and 1.12, given the corresponding SNRs of 32 and 27. Thus, we infer that the radio emission observed by VLA in each frequency band is resolved, albeit only slightly.

\section{Discussion}\label{sec:diss}

\subsection{Origin of the compact radio emission}
The interpretation of radio spectra in AGNs depends on various contributing components, which complicates identifying the dominant emission mechanism. For powerful radio-loud AGNs, a flat radio spectrum is typically considered to be dominated by the central compact radio core, while a steep radio spectrum indicates dominance by extended structures such as jets or lobes~\citep{Condon.92.araa}. 
In radio-quiet AGNs, however, the situation is more complex, as the detected radio emission can originate from a variety of mechanisms, including star formation, AGN-driven winds, free-free emission from photoionized gas, low-power jets, and coronal activity from the innermost accretion disk~\citep{Panessa.19.na}. Although we have some understanding of the radiative mechanisms of AGNs, the radio emission processes of low-mass AGNs are still poorly understood. A statistical analysis by \citet{wu.24.apjs} reveals an unexpectedly high radio-loud fraction ($>$60\%) in their low-mass AGN sample, suggesting that substantial radio emissions may originate from star formation within their host galaxies or that the radio loudness is biased by the low luminosities characteristic of these low-mass AGN.
Therefore, it is essential to consider the contributions from the host galaxies thoroughly before making definitive conclusions.

\subsubsection{Confirmation of AGN activity}

To rigorously evaluate potential star formation contribution to the observed radio emission, we performed a comprehensive multi-band analysis:

\textbf{{Radio-based SFR estimates:}}
We \sout{also} calculate the expected monochromatic luminosity at 1.4 GHz based on the estimator from~\citet{Kennicutt.12.araa}: 
\begin{equation}\label{eqn:1}
\log L_{1.4 \mathrm{GHz}}\left(\mathrm{erg} \mathrm{s}^{-1} \mathrm{~Hz}^{-1}\right) = \log \mathrm{SFR}\left(\mathrm{M}_{\odot} \mathrm{yr}^{-1}\right) + 28.2, 
\end{equation}
which also incorporates the star formation rate (SFR) calibration from~\citet{Murphy.11.apj}.
The SFR = $0.15 \pm 0.01\ \mathrm{M}_{\odot}\ \mathrm{yr}^{-1}$ derived from UV--IR SED fitting \citep{Kawamuro.24.apj}.
This results in an expected flux density of $222 \pm 14.8\ \mu$Jy at 1.4 GHz given the SED-based SFR. By extrapolating this flux density to the 2–10 GHz range using a power-law slope of $\alpha = 0.7$, a typical value for star formation-driven emissions \citep{Condon.92.araa, Panessa.19.na}, we find that the expected radio emission from star formation is significantly lower than the observed flux densities in both the ATCA and VLA data in Figure~\ref{fig:spectral}.
 
\textbf{{Radio spectral properties:}}
While the measured spectral index ($\alpha \sim 0.63$) is similar to that expected from star formation ($\alpha \sim 0.7$) \citep{Condon.92.araa}, the compact morphology ($<1\ \mathrm{kpc}$) differs from the extended radio emission typically associated with star formation \citep[on several kpc scales; see][]{Murphy.12.apj}.

\textbf{{Multi-band diagnostics:}}
(a) The Baldwin, Phillips, and Terlevich (BPT) diagnostic ratios 
$\mathrm{[O\ III]}/{\mathrm{H}\beta} \quad \text{and} \quad \mathrm{[N\ II]}/{\mathrm{H}\alpha}$ place POX 52 firmly in the AGN region \citep{ludwig.12.apj}.
(b) The mid-IR colors from Wide-field Infrared Survey Explorer (WISE) ($W_1 - W_2 = 0.86$) are consistent with AGN emission \citep{Stern.12.apj}.
(c) The observed X-ray luminosity ($L_{\mathrm{X}} \sim 10^{42}$ erg s$^{-1}$) is two orders of magnitude higher than expected from star formation alone \citep[$L_{\mathrm{X,SF}} \approx 10^{40}$ erg s$^{-1}$ for SFR $= 0.15\ M_{\odot}$ yr$^{-1}$, following][]{Lehmer.16.apj}.

\textbf{{Host galaxy properties:}}
Hubble Space Telescope (HST) imaging reveals POX 52 as a dwarf elliptical galaxy lacking spiral arms or prominent star-forming regions~\citep{thornton.08.apj}. The stellar population analysis indicates a predominantly old stellar population with mean age $>$5 Gyr~\citep{barth.04.apj}.

\textbf{{Temporal characteristics:}}
The observed radio variability on monthly timescales (Section 4.1.3) cannot be explained by star formation~\citep{wang.23b.mn}, which typically varies on much longer timescales
\citep[$>$100 yr;][]{Condon.92.araa}.

Therefore, based on our quantitative analysis of the star-forming contributions and above multi-band evidence, we conclude that the observed radio emission in POX~52 predominantly originates from AGN activity.

\subsubsection{The origin of the AGN activity}
Although we have ruled out star formation as the origin of the radio emission, there are still multiple mechanisms by which AGNs could generate this emission, including coronal activity, winds/outflows, and jets. Distinguishing among these possibilities is very challenging given the limited resolution.

While the ratio of $L_\mathrm{R}$ to $L_\mathrm{X}$ appears consistent with the G\"udel-Benz relation~\citep{Guedel.93.apjl} for coronal emission, two factors argue against a coronal origin. 
(a) The coronal component originates from the innermost region surrounding the black hole, spanning a range of approximately 10–10,000 $R_\mathrm{g}$~\citep[Schwarzschild radius; equivalent to $10^{-7}$–$10^{-4}$ pc for a $10^5 M_\odot$ black hole,][]{Panessa.19.na}. In contrast, our observed radio emission is resolved on a scale of 100~pc in VLA observations, which is orders of magnitude larger than the typical size of a corona.
(b) The steep spectrum is inconsistent with the flat or inverted spectra that are typically associated with corona~\citep{Raginski.16.mn}.

The steep spectrum component observed could be attributed to optically thin synchrotron emission, potentially dominated by plasma blobs within a sub-relativistic jet or an AGN wind/outflow \citep{Laor.19.mn}. In the observations of AGNs hosting SMBHs, jets typically extend over hundreds of parsecs, whereas outflows can span thousands of parsecs \citep{Panessa.22.mn}. However, these structures tend to be more compact in the IMBH regime. Thus, it remains challenging to definitively conclude whether the observed radio emission originates from a jet, an outflow, or a combination of both, although the elliptical components in VLA maps correspond to a physical scale of several hundred parsecs (see~Table~\ref{table:info_obs}). 

\subsubsection{Radio variability}
As shown in Figure \ref{fig:spectral}, the observed radio emission detected by ATCA are higher than that in VLA in the overlapping frequency. This discrepancy could be due to the VLA's higher angular resolution resolving out diffuse or extended emission detected by ATCA, or it might indicate real variability in the radio emission.

To verify the resolution scenario, we employed ten short-baseline telescopes from the VLA to conduct mapping and model fitting. In this configuration, the major and minor axes of the synthesized beam are larger than those obtained with ATCA, namely a lower angular resolution compared to that of ATCA. Consequently, the VLA should have captured the diffuse or extended emission observed by ATCA. However, we found that the peak intensity and integrated flux densities at 5~GHz ($f_{\rm peak}$ = 138 $\pm$ 12~$\mu$Jy/beam and $f_{\rm int}$ = 158 $\pm$ 26~$\mu$Jy) are still lower than those measured by ATCA in Table \ref{table:info_obs}, indicative of the short-term AGN variability over the 8 months between ACTA and VLA observations. Moreover, an upper limit of 78~$\mu$Jy was reported in VLA's A configuration C-band observations in 2004 \citep{thornton.08.apj}, further confirming long-term radio variability over two decades.

\subsection{The radio detections in low-mass AGNs}

Radio detections of low-mass AGNs remain rare due to their faint and weak radio emission. A well-studied case is NGC 4395, with the angular resolution from sub-arcsec~(e.g., $0\farcs29\times0\farcs26$) to mas-scales~(e.g., 3.13$\times$0.95~mas)~\citep{Saikia18, yangjun.22.mn}. 
VLA observations at 12--18 GHz~(beam: $0\farcs13\times0\farcs12$) reveal an elongated structure with a total flux density of approximately 370 $\mu$Jy,
however, it turns out to be nuclear shocks likely formed by the IMBH's episodic ejection or wide-angle outflow rather than the core-jet in the mas-scale~\citep{yangjun.22.mn, yangjun.23.mn}. 
Extending to other low-mass AGNs with pointing observations, most of sub-arcsec scale deep observations show ``compact" structure, namely single component source, unresolved or slightly resolved ~\citep{Gultekin.14.apjl,gultekin.22.mn,paul.24.apj}. 

Leveraging comprehensive wide-field radio survey, \citet{qian.18.apj} and \citet{wu.24.apjs} successfully identified radio counterparts for 151 optically selected low-mass AGNs, cataloged as 132 compact and 19 extended sources, based on previous low-mass AGN catalog \citep{dong.12.apj,liu.18.apjs}. This identification was achieved by integrating data from Square Kilometer Array (SKA) pathfinders such as the LOFAR Two-meter Sky Survey \citep[LoTSS,][]{shimwell.17.aa, shimwell.19.aa, shimwell.22.aa}, the Apertif-shallow survey (Adams et al. 2024 in prep.), and the VLASS epoch 1 Quick Look, as well as through deep mining of data from the Faint Images of the Radio Sky at Twenty-Centimeter (FIRST) survey. This effort has quadrupled the number of known low-mass AGNs with radio counterparts, thereby establishing a solid foundation for future high-frequency observational studies.

\subsection{The Fundamental Plane of Black Hole Activity}

Accretion, outflows, and jets are prevalent across a wide range of astrophysical systems, varying in size, mass scale, and environmental conditions \citep{Connors.17.phdthesis}. Over the past few decades, we have learned that black hole accretion seems to be a scale-invariant process from X-ray binaries to SMBH, and thus establishing the empirical relation linking an accreting BH's mass (${M_{\rm BH}}$), radio luminosity ($L_{\mathrm{R}}$), and X-ray luminosity ($L_{\mathrm{X}}$), known as the fundamental plane of BH activity \citep{Merloni.03.mn, Falcke.04.aa, Gultekin.09.apj, yuanfeng.09.apj, Plotkin.12.mn, Gultekin.14.apjl, chang.21.mn, wangyj.24.aa}. This relationship has been thoroughly explored from the stellar-mass to the SMBH regimes, significantly enhancing our understanding of the underlying physics of black hole accretion and jet dynamics. However, the application of this fundamental plane to IMBHs remains largely unexplored \citep{Maccarone.04.mn, gultekin.22.mn}, and it is still unclear whether IMBHs conform to this empirical relation.

\begin{figure}
\hspace{-1cm}
    \includegraphics[width=9.5cm]{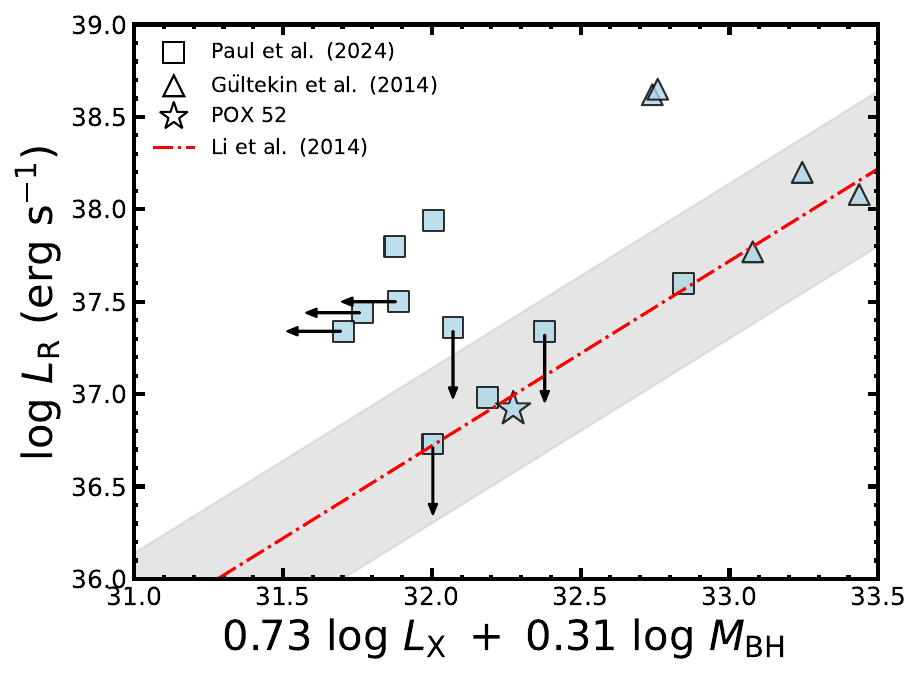}
    \caption{
    The IMBH candidates in the fundamental plane of BH activity.
The red dashed-dotted line represents the best fit for a sample of 227 high Eddington ratio ($\overline{\lambda}_{\mathrm{Edd}} = 0.3$) radio-quiet broad-line AGNs \citep{lizhaoyu.08.apj}, with the shaded area indicating the 1$\sigma$ uncertainty (0.42 dex) in the $L_\mathrm{R}$ direction. Squares and triangles denote 15 radio-quiet, low-mass AGNs with Eddington ratios greater than 0.1 and black hole masses in the range $10^{5.5} M_{\odot} < M_{\mathrm{BH}} < 10^{6.3} M_{\odot}$. For this comparison sample, we obtained values for the 5 GHz radio luminosity, 2–10 keV X-ray flux, black hole mass, and Eddington ratio from the literature~\citep{green.07.apj,Gultekin.14.apjl,paul.24.apj}.}
    \label{fig:fp}
\end{figure}

Observational evidence indicates that radio-quiet and radio-loud AGNs have distinct fundamental planes~\citep{wangran.06.apj, lizhaoyu.08.apj,Bariuan.22.mn}. Moreover, for both types of AGNs, the fundamental plane shows a significant dependence on the Eddington ratio~\citep[as summarized in][]{wangyj.24.aa}. Thanks to the X-ray data and deep radio observations~\citep{Gultekin.14.apjl,paul.24.apj} of 15 high-Eddington ratios ($0.1<\lambda_{\mathrm{Edd}}<1$; $\left\langle\lambda_{\text {Edd }}\right\rangle=0.37$), radio low-mass AGNs, we can extend the fundamental plane to the IMBH regmin, together with POX 52 ~\citep[$\lambda_{\mathrm{Edd}}\sim 0.5$,][]{barth.04.apj,thornton.08.apj}.

Given the limited size of our sample, the considerable uncertainties of black hole masses, and the need for refined radio and X-ray luminosity measurements, we avoid to derive new relationships in our work. Instead, we simply check for consistency with an existing empirical relation described by \citet{lizhaoyu.08.apj}, which was established for a sample of 227 broad-line AGNs. Similar to our sample, this AGN sample is radio-quiet with an average Eddington ratio of 0.3.

Our findings, illustrated in Figure~\ref{fig:fp}, demonstrate that POX 52, along with approximately half of the low-mass radio-quiet AGNs, follows the fundamental plane of supermassive AGNs as detailed in \citet{lizhaoyu.08.apj}, which share similar Eddington ratios ranging from 0.1 to 1, albeit with some outliers.

These deviations could stem from several factors: potentially overestimated radio emissions in low-mass AGNs due to additional contributions such as star formation \citep{yangxl23}; underestimations of black hole mass, considering a $\sim$0.5 dex uncertainty based on the single-epoch spectrum \citep{Shen13}; the non-simultaneous multi-band measurements; or intrinsic discrepancies. The fundamental plane sheds important light on the accretion physics and X-ray emission origins of central engines, our results suggest that the accretion and jet processes may be comparable across accreting systems with different black hole masses, given a similar Eddington ratio, within the radio-quiet AGN population.             

\section{Conclusions} \label{sec:con}

Our discovery of radio emission from POX 52 represents a significant progress in the study of IMBHs and AGN physics. This first successful radio detection of one of the most promising nearby IMBH candidates provides crucial evidence for AGN activity in the IMBH regime, while demonstrating the feasibility of detecting radio emission from high-Eddington ratio IMBHs. The detection enables direct comparison with SMBH systems, bridging a critical gap in our understanding of black hole physics across the mass scale. The main conclusions are as follows:
\begin{enumerate}
\item
The detailed characterization of POX 52's radio properties has revealed several important physical insights. The steep spectrum (VLA: $\alpha = 0.63 \pm 0.03$; ATCA: $\alpha = 0.99 \pm 0.26$) clearly indicates the presence of optically thin synchrotron emission, while the observed radio variability on monthly timescales provides new constraints on emission mechanisms. 
The deconvolved angular sizes of the elliptical Gaussian components fitted in the VLA \textsc{\footnotesize CLEAN} maps favor the low-power jet or radiation-driven wind/outflow as the origin of the radio emission.
\item
POX 52, along with several other objects, aligns well with the low-mass extension of the fundamental plane for high-Eddington ratio, radio-quiet AGNs \citep{lizhaoyu.08.apj}. This extension places critical new constraints on black hole scaling relations and reinforces the paradigm of self-similar accretion processes across seven orders of magnitude in black hole mass. However, approximately half of the other low-mass AGNs with high Eddington ratios deviate from this relationship, emphasizing the need to investigate intrinsic physical processes or potential errors in the measurement of fundamental plane variables.
\end{enumerate}

This study successfully detects radio emission from the IMBH in POX 52 at sub-arcsec scales, establishing a critical stepping stone for future investigations at milli-arcsec resolutions. The complementary use of deep, multi-frequency radio imaging with ATCA and VLA has provided valuable insights into the characterization of IMBH properties, offering a solid foundation for exploring similar objects in future studies.

These results can provide a crucial link to broader astrophysical questions and hold significant implications for understanding black hole growth and evolution across cosmic time. By characterizing the properties of nearby IMBHs such as POX 52, we uncover critical insights into the origins of black hole seeds in the early universe—a subject of heightened importance in light of recent JWST discoveries of luminous high-redshift quasars.
Looking ahead, upcoming facilities—including the next-generation Very Large Array (ngVLA; \citealt{Murphy.18.aspcs}), the Square Kilometer Array Observatory (SKA; \citealt{braun.19.arxiv}), and the FAST Core Array (\citealt{Jiang24})—are poised to play a pivotal role in advancing sensitivity and resolution, enabling transformative breakthroughs in the study of IMBHs and their cosmological implications.

\label{conclu}

\begin{acknowledgements}

We thank the anonymous referee for helpful comments. This work is supported by the National Key R\&D Program of China No. 2022YFF0503402, 2023YFA1607903, and Future Network Partner Program, CAS, No. 018GJHZ2022029FN, Overseas Center Platform Projects, CAS, No. 178GJHZ2023184MI, National Natural Science Foundation of China (NSFC) No. 1247030223. MFG is supported by the National Science Foundation of China (grant 12473019), the National SKA Program of China (Grant No. 2022SKA0120102), the Shanghai Pilot Program for Basic Research-Chinese Academy of Science, Shanghai Branch (JCYJ-SHFY-2021-013), and the China Manned Space Project with No. CMSCSST-2021-A06.
QY and MZ are supported by the National Key R\&D Intergovernmental Cooperation Program of China (2022YFE0133700), the National Natural Science Foundation of China (12173078).
RZS acknowledges the support from the China Postdoctoral Science Foundation (Grant No. 2024M752979).
NC acknowledges the support from the Xinjiang Tianchi Talent Program.
This work has made use of data from the European Space Agency (ESA) mission Gaia (\url{https://www.cosmos.esa.int/gaia}) processed by the Gaia Data Processing and Analysis Consortium (DPAC, \url{https://www.cosmos.esa.int/web/gaia/dpac/consortium}).
The Australia Telescope Compact Array is part of the Australia Telescope National Facility (grid.421683.a) which is funded by the Australian Government for operation as a National Facility managed by CSIRO. We acknowledge the Gomeroi people as the traditional owners of the Observatory site.
The National Radio Astronomy Observatory (NRAO) is a facility of the National Science Foundation operated under cooperative agreement by Associated Universities, Inc. This paper makes use of the VLA data from program 23A-163.

\end{acknowledgements}

\software{
\textsc{\footnotesize{CASA~\citep{CASA.22.pasp}}}, RACS-tool\footnote{https://github.com/alecthomson/RACS-tool}.
}

\bibliography{main}{}
\bibliographystyle{aasjournal}

\end{document}